\documentclass[aps,prl,twocolumn,showpacs,superscriptaddress]{revtex4}

\newfont{\BBB}{msbm10 scaled\magstephalf}

\newcommand{\BBC}{\mbox{\BBB C}}

\begin{document}
\title{Broken quantum symmetry and confinement phases in planar physics}
\author{ F.~A.~Bais}
\affiliation{
Institute for Theoretical Physics,
Valckenierstraat 65, 1018 XE Amsterdam, The Netherlands}
\author{B.~J.~Schroers}
\affiliation{Department of Mathematics, Heriot-Watt University,
Edinburgh EH14 4AS, United Kingdom}
\author{J.~K.~Slingerland}
\affiliation{
Institute for Theoretical Physics,
Valckenierstraat 65, 1018 XE Amsterdam, The Netherlands}

\date{May 5, 2002}

\begin{abstract}
\noindent 
Many two-dimensional physical systems have symmetries which are
mathematically described by quantum groups (quasi-triangular Hopf
algebras). In this letter we introduce the concept of a spontaneously
broken Hopf symmetry and show that it provides an effective tool for
analysing a wide variety of phases exhibiting many distinct confinement
phenomena.
\end{abstract}
\maketitle


\section{Introduction}

Planar quantum physics is known to exhibit many surprising properties
like charge fractionalization, spin-charge separation, fractional and
non-abelian statistics. Important analogies show up between apparently
different systems like fractional quantum Hall systems and rotating
bose condensates.  Many of the special features are based on a subtle
interplay between particles and their duals, e.g. between charges and
fluxes, spins and vortices, particles and quasiparticles or
defects. These features are often a consequence of topological
interactions between the relevant degrees of freedom. From a
mathematical point of view many of these aspects are related to
nontrivial realizations of the braid group.  The appearance of the
braid group is linked quite generically to the presence of an
underlying quantum symmetry described by a (quasi-triangular) Hopf
algebra.  Quantum groups naturally provide a framework in which
abelian or non-abelian representations of the braid group can be
constructed explicitly.  Moreover, particles and their duals are
treated on equal footing in this framework.  As a result it is
possible to give a systematic and detailed description of the spin and
statistics properties of the relevant degrees of freedom.

The generic appearance and therefore importance of Hopf symmetries in
two-dimensional systems provide a strong physical motivation for
studying what happens to such systems if one of the (bosonic) fields
acquires a vacuum expectation value which breaks the Hopf
symmetry. How does a phase with broken Hopf symmetry manifest itself
physically and how one can such phases be characterized?  In the case
of breaking of ordinary gauge symmetries one usually finds that masses
for vector particles are generated and/or massless scalars show up.  A
further - and equally important - aspect of symmetry breaking is the
impact on topological defects: some of them will disappear from the
spectrum and new ones may show up depending on the properties of the
order parameter \cite{Bais:1981nn}.  As we will show this is only the
simplest case, with other and more complicated situations arising when
dual (dis)order parameters -- for example, the density of magnetic
vortices -- come into play \cite{mandelconf,'tHooft:1977jd,hooftconf}.

In this paper we report on general results from the study of
(dis)order parameters that carry representation labels of a Hopf
algebra $A$. The (dis)order parameter breaks the Hopf symmetry to some
Hopf subalgebra $T$. The analysis shows that the representations of
$T$ fall into two sets. One contains representations that get confined
in the broken phase, while the other contains non-confined
representations.  The tensor products of $T$ representations allow one
to determine the ``hadronic'' composites that are not confined.  The
non-confined representations together form the representation ring of
a smaller algebra $U$, which is the residual symmetry of the the
effective low energy theory of non-confined degrees of freedom. We
find that both confined and not confined representations can be
electric, magnetic or dyonic in nature, depending on the type of
(dis)order parameter one assumes is condensed. We have relegated an
extensive mathematical treatment of these problems to a separate
paper \cite{baslsch}, to which we refer the reader for more detailed
statements and proofs.

\section{Hopf symmetry}

In this section we briefly summarize some essential properties of a
Hopf algebra \cite{chapress}, choosing a relative simple class as an
example. This class describes the symmetry that arises if one breaks
for example a non-abelian continuous group $G$ to a discrete subgroup
$H$, giving rise to what is known as a discrete gauge
theory \cite{Bais:1980vd,krausswil,preskrauss,dgt5}. Such a model
contains magnetic defects which carry a flux labeled by a group
element of $H$.  The group $H$ acts on fluxes by conjugation, so that
fluxes in the same conjugacy classes form irreducible multiplets.  If
the group is non-abelian one finds that the fluxes, when parallel
transported around each other, generate non-abelian Ahoronov-Bohm
phases. The underlying Hopf symmetry in this case turns out to be the
quantum double $ A = D(H)$ of the group $H$ \cite{dpr,dgt1}. This
double has more structure than the group $H$ because $D(H)\equiv
F(H)\tilde \otimes \BBC H$.  Here $F(H)$ are the functions on the group
and $\BBC H$ is the group algebra of $H$ (the linear span of group
elements with the given group product). The symbol $\tilde \otimes$
indicates that $D(H)$ is the tensor product of $F(H)$ and $\BBC H$ but
that the multiplication of two elements of $D(H)$ is
``twisted''. Explicitly, the multiplication rule for two elements in
$D(H)$ is :
\begin{equation}
(f_1\otimes h_1) (f_2\otimes h_2)(x) = f_1(x)f_2(h_1xh_1^{-1})
\otimes h_1h_2  \;\;\;\; x\in H
\end{equation}
Note that the product in the $\BBC H$-component is the ordinary group
multiplication but that the pointwise multiplication of functions is
twisted by the conjugation action of $H$. Physically, we think of $H$
as the ``electric'' gauge group generated by $\{1\otimes h\}$ while
the $F(H)$ component is a ``magnetic symmetry'' generated by
$\{f\otimes e\}$. The unitary irreducible representations of $D(H)$
are denoted by $\Pi_\alpha^A$. Here $A$ is a a magnetic (flux) quantum
number labeling a conjugacy class of $H$ and $\alpha$ is an
electric quantum number labeling a representation $\alpha$ of the
centralizer $N_A$ of that conjugacy class.
We see that the trivial class $\{e\}$ (consisting of the unit element
of $H$) gives the usual representations of $H=N_{\{e\}}$ corresponding
to the purely electric states.  Conversely the representations with
the trivial $\alpha$ representations are the purely magnetic
multiplets.  At this point one should observe that the labeling of
the dyonic (i.e. mixed) states already takes care of a well known
subtlety, namely the obstruction to defining full $H$-representations
in the presence of a non-abelian magnetic flux. D(H) has a trivial
representation $\varepsilon$ (the co-unit) defined by $
\varepsilon(f\otimes h) = f(e)$.  There is a canonical way in which
tensor product representations are defined, leading to a Clebsch
Gordon series:
\begin{equation}
\Pi_\alpha^A \otimes \Pi_\beta^B \cong N_{\alpha\beta C}^{A B
\gamma} \Pi_\gamma^C.
\end{equation}
The final ingredient is the R-matrix $R \in D(H)\otimes D(H)$
implementing the braid operation on a two particle state through
\begin{equation}
{\cal R} \equiv \sigma \cdot (\Pi_\alpha^A \otimes \Pi_\beta^B) (R),
\end{equation}
where $\sigma$ is the ``flip'' operation, interchanging the order of the
factors in the tensor product. The ${\cal R}^2$ operator yields the
monodromy, or generalized Ahoronov Bohm phase factor. 
 
We note that Hopf symmetry plays a role in all planar systems that
have a conformal field theory description, such as 2 dimensional
critical phenomena, fractional quantum Hall states \cite{MR,SB}, and
the world sheet picture of string theory. In these systems the tensor
product rules of the quantum group are directly related to the fusion
rules of the chiral algebra of the conformal field theory. The
(quasi)particle excitations carry representations of that quantum
group and the same mathematical tools can be used to characterize the
Hall plateau states and their excitations.

\section{Hopf symmetry breaking}

Let us imagine a condensate forming in a state $\mid v \rangle$ in the
carrier space of some representation $\Pi^A_\alpha$.  Then we may
define {\em the maximal Hopf-subalgebra} $T$ of $A$ which leaves $\mid
v \rangle$ invariant. Explicitly, elements $P$ of $T$ satisfy
\begin{equation}
\Pi^A_\alpha(P)\mid v \rangle =
\varepsilon(P)\mid v \rangle \qquad\forall P \in T.
\end{equation}
Given the original Algebra $A$ there is a systematic way of
calculating $T$.  The most familiar example is the case where $A$ is
the group algebra of an ordinary group $H$. In that case one easily
checks $T$ is the group algebra of a subgroup of $H$, thus reproducing
the well-known form of symmetry breaking.  A first nontrivial case is
$A = F(H)$.  In that case the algebra of functions on the group $H$
gets broken to the algebra of functions on the quotient group $H/K$,
where $K$ is some normal subgroup of $H$ (i.e. $HKH^{-1}=K$).

Let us now take a closer look at the situation for $A=D(H)$.  If we
break by a purely electric condensate $\mid v \rangle \in V^e_\alpha$
then the magnetic symmetry is unbroken but the electric symmetry $\BBC
H$ is broken to $\BBC N_v$, with $N_v \subset H$ the stabilizer of
$\mid v \rangle$.  In that case we get $T=F(H)
\;\;\tilde{\otimes}\;\;\BBC N_v$.

We may also break by a gauge invariant purely magnetic
state. Interestingly enough one such a state exists for each conjugacy
class and corresponds to an unweighted sum of the basis vectors
representing the group elements in the class: $\mid v \rangle =
\sum_{a\in A} \mid a\rangle \in V^A_1$. The group action of $H$ leaves
this state invariant:
\begin{equation}
\Pi^A_1(1
\otimes h)\mid v \rangle = \sum_{h\in A} |hah^{-1}> =
\sum_{a\in A} \mid a\rangle = \mid v \rangle.
\end{equation} 
In this case one may show that the unbroken Hopf algebra is $T =
F(H/K_A)\tilde{\otimes}\BBC H$ with $K_A \subset H$ the subgroup
generated by the elements of class $A$. This reduction of the symmetry
reflects the physical fact that the fluxes can only be defined up to
fusion with the fluxes in the condensate.

As a final example we consider what happens if the condensate
corresponds to a single flux state $\mid v \rangle=\mid g \rangle$
with ($g \in A$). Now one finds $T=F(H/K_A)\tilde{\otimes}\BBC N_g$
with $N_g=\{h\in H \mid hg=gh\}$, showing that both electric and
magnetic symmetry are partially broken.

\section{Confinement}

Consider now the physical situation after the breaking has taken
place. As the ground state has changed we should discuss the fate of
the (quasi)particle states belonging to the representations of the
residual Hopf algebra $T$. These representations can be
constructed\footnote{This is hard in general but rather
straightforward in the not uncommon case where $T$ is a so called
transformation group algebra \cite{baslsch}} and describe the
excitations in the broken phase.  Furthermore, there is a
decomposition of representations of the algebra $A$ into
representations $\Omega_j$ of $T\subset A$. Now it may happen that the
braiding of the condensed state $|v\rangle \in V_0$ and some
(quasi)particle state $|p\rangle $ in a representation $\Omega_j$ is
nontrivial.  If this happens, the vacuum state is no longer single
valued when transported around the (quasi)particle. Consequently the
new ground state does not support a localized excitation of the type
$\Omega_j$ and will force it to develop a string-like singularity,
i.e. a domain wall ending on it. Such a wall carries a constant energy
per unit length and therefore the particle of type $\Omega_j$ will be
confined. The upshot is that we can use braid relations of the $T$
representations $\Omega_j$ with the ground state representation
$\Pi_0$ of $A$ to determine whether or not the corresponding particles
are confined. Physically speaking this procedure is like imposing a
generalized Dirac charge quantization condition to determine the
allowed non-confined excitations in a given phase. In general the
determination of these braid relations of the $T$ and $A$
representations is a difficult problem. For detailed calculations we
refer to our paper \cite{baslsch}. It is also shown there that all $T$
representations which have trivial braiding with the vacuum
representation can survive as localized states in the broken phase.

Consistency requires that the non-confined representations should form
a closed subset under the tensor product for representations of $T$.
One may show that this is the case and that the subset of non-confined
representations can in fact be viewed as the representations of yet
another Hopf algebra $U$. Mathematically, $U$ is the image of a
surjective Hopf map
\begin{equation} 
\label{sur}
\Gamma:T\rightarrow U.
\end{equation}
The $U$ symmetry characterizes the particle like representations of
the broken phase. Under quite general circumstances $U$ itself is
again quasi-triangular, implying that it features an R-matrix which
governs the braid statistics properties of the non-confined
excitations in the broken phase.  Returning to $T$, it is clear that
the tensor product rules for confined $T$ representations allow one to
construct multi-particle composite ("hadronic") states which belong to
non-confined representations.

For a complete characterization of the excitations in the broken phase
we should comment on the strings attached to confined
particles. These are not uniquely characterized by their endpoints
because one can always fuse with non-confined particles.  It turns
out that the appropriate mathematical object characterizing the
strings is the Hopf kernel ${\rm Ker}(\Gamma)$ of the Hopf map
(\ref{sur}).

To illustrate these concepts we return to the examples mentioned in
the previous Section. The first example concerned a purely electric condensate
which just breaks the electric gauge group to $N_v$ so that, as
mentioned, $T\equiv F(H)\tilde{\otimes}\BBC N_v$. One
obtains $U \equiv F(N_v)\tilde{\otimes}\BBC N_v\equiv
D(N_v)$ and ${\rm Ker}(\Gamma)=F(H/N_v)$.  Physically, this means
that the only surviving representations are those which have magnetic
fluxes corresponding to elements of $N_v$ while the states with fluxes
in the set $H-N_v$ get confined. In short, partial electric breaking
leads to a partial magnetic confinement. The distinct walls are now in
one-to-one correspondence with the $N_v$ cosets in $H-N_v$.

The second example had the gauge invariant magnetic condensate and
we found that $T=F(H/K_A)\tilde{\otimes}\BBC H$ with $K_A
\subset H$ the subgroup generated by the elements of class $A$. In
this case we find that $ U=D(H/K_A)$ with ${\rm Ker}(\Gamma)=\BBC
K_A$.  Thus, only electric representations which are $K_A$ singlets
survive while the others get confined. Partial or complete magnetic
breaking will result in partial or complete electric confinement,
depending on $K_A$. The walls in this phase are labeled by the
representations of $K_A$.

Finally the pure flux condensate $|g\rangle$, which has
$T=F(H/K_A)\tilde{\otimes}\BBC N_g$, leads to a phase for which
$U=D(N_g/K_A\cap N_g)$ and ${\rm
Ker}(\Gamma)=F((H/K_A)/\bar{N}_g)\tilde\otimes\BBC (K_A\cap
N_g)$. Here $\bar{N}_g$ is the subgroup of $H/K_A$ which consists of
the classes $nK_A$ with $n\in N_g$. In this case we have a breaking of
magnetic and electric symmetry leading to a (partial) confinement of
both. We do not discuss dyonic condensates here, not because of
essential complications but rather because of notational
inconveniences. The same analysis can be applied.

\section{Conclusion}
Field theories on a plane may contain (quasi-)particles with
nontrivial topological interactions and braid statistics.  Such
systems often have a hidden quantum symmetry described by a Hopf
algebra $A$. Representations of such algebras have the attractive
feature that they treat ordinary and topological quantum numbers on
equal footing.

In this letter we investigated what happens when such a Hopf symmetry
$A$ gets broken to a Hopf algebra $T$ by a vacuum expectation value of
some field carrying a representation of $A$. We showed that
generically there is a hierarchy of three Hopf algebras $A$, $ T$ and
$ U$ which play a role in this situation.  The representations of $T$
fall into two sets, one set being confined while the other is not. The
latter can be interpreted as the representations of the Hopf
subalgebra $U$ which is the residual symmetry of the broken phase.
The tensor product rules of $T$ representations tell us also what the
non-confined composites (i.e. the ``hadronic'' excitations) will be.
The tools and methods described here enable one to analyze a wide
variety of phases, each with its specific pattern of (partial)
confinement properties, and the way these phases are linked.

\bibliographystyle{unsrt}

\end{document}